\documentclass[journal]{IEEEtran}
\usepackage{comment}

\ifCLASSINFOpdf
   \usepackage[pdftex]{graphicx}
\else
\fi
  \usepackage[caption=false,font=normalsize,labelfont=sf,textfont=sf]{subfig}
\usepackage{multirow}
\usepackage{hhline}
\usepackage{pgf-pie}
\usepackage{url}
\begin{document}
% \title{AudioSentibank: Large-scale Vocabulary for detecting sentiment and semantics in audio content}
%\title{AudioSentibank: Large-scale Semantic Ontology of Acoustic Concepts for Audio Content Analysis}
\title{AudioPairBank: Towards A Large-Scale Tag-Pair-Based Audio Content Analysis}

\author{\IEEEauthorblockN{Sebastian S\"ager}
\IEEEauthorblockA{University of Kaiserslautern\\
DFKI\\
67663 Kaiserslautern, Germany\\
Email: s\_saeger13cs.uni-kl.de}
\and
\IEEEauthorblockN{Benjamin Elizalde}
\IEEEauthorblockA{Carnegie Mellon University\\
5000 Forbes Ave, Pittsburgh, PA 15213, USA\\
Email: bmartin1@andrew.cmu.edu}
\and
\IEEEauthorblockN{Damian Borth}
\IEEEauthorblockA{DFKI\\
67663 Kaiserslautern, Germany\\
Email: damian.borth@dfki.de}}

% suggestion - let's discuss this to get the backstrory from Prof. Dengel, the advisor of Sebastian
\author{Sebastian~S\"ager,
		and~Benjamin~Elizalde,
        and~Damian Borth~\IEEEmembership{IEEE Member},
        and~Christian Schulze,
        and~Bhiksha~Raj~\IEEEmembership{IEEE Fellow},
        and~Ian~Lane~\IEEEmembership{IEEE Member}
        %,
        %and~Andreas~Dengel% <-this % stops a space
}


\markboth{IEEE/ACM Transactions on Audio, Speech, and Language Processing, July~2017}%
{Shell \MakeLowercase{\textit{et al.}}: Bare Demo of IEEEtran.cls for IEEE Journals}



\maketitle

% in the abstract or keywords. in order to do our thing is was first necessary to collect. with it's implications, previously. 
%benchmark-correlation possible to do such classification
%first contributions collecting dataset implications given that didnt exist.
\begin{abstract}
Recently, sound recognition has been used to identify sounds, such as \textit{car} and \textit{river}. However, sounds have nuances that may be better described by adjective-noun pairs such as \textit{slow car}, and verb-noun pairs such as \textit{flying insects}, which are under explored. Therefore, in this work we investigate the relation between audio content and both adjective-noun pairs and verb-noun pairs. Due to the lack of datasets with these kinds of annotations, we collected and processed the AudioPairBank corpus consisting of a combined total of 1,123 pairs and over 33,000 audio files. One contribution is the previously unavailable documentation of the challenges and implications of collecting audio recordings with these type of labels. A second contribution is to show the degree of correlation between the audio content and the labels through sound recognition experiments, which yielded results of 70\% accuracy, hence also providing a performance benchmark. The results and study in this paper encourage further exploration of the nuances in audio and are meant to complement similar research performed on images and text in multimedia analysis.  

\end{abstract}

%Webiste for Keywords:
%http://www.ieee.org/documents/taxonomy_v101.pdf

% \fi
\IEEEpeerreviewmaketitle

\begin{comment}

\section{Index - this section will be deleted}
\textbf{+Intro-Related work}\\
-Intro to audio\\
-Audio applications--evaluations, datasets\\
-Subjectivity/Sentiment-- in speech, music and images, and sounds\\
-Our contribution\\
\textbf{+Collection of dataset}\\
-Overview\\
-Selection of classes: A/VNPs \\
-Download of audio and source\\
-Analysis of Audio SentiBank\\
==Tags co-occurrence\\
==Number of Tags\\
==Duration of audio\\
==Audio samples per class\\
==Users distribution\\
-Filtering of data\\
==Sampling below 16kHz\\
==Files with no tags\\
==Tags with loop and processed and music***\\
==Duration outliers and very long\\
==Small amount of samples\\
==Implausible pairs\\
==Dominant user\\
-Final dataset: num. classes, examples\\
\textbf{+Experiments}\\
-Binary classification setup\\
-Multiclass classification setup\\
\textbf{+Results and Analysis of Qualifiers}\\
-Overall benchmark for binary and multiclass with metrics: Acc, F1score, etc…\\
-Performance of  top/low ANPs and VNPs;  ANPs vs VNPs?\\
-Overall Consusability based on Confusion matrix\\
-Similar A different N, confusability and performance\\
-Similar V different N, confusability and performance\\
-Similar N different A/V, confusability and performance\\
\textbf{+Conclusions}\\
\end{comment}

%%%%%%%%%%%%%%%%%%%%%%%%%%%%%%%%%%%%%%%%%%%%%%%%%%%%%%%%%%%%%%
%
%	Introduction & Related
%
%
%
%%%%%%%%%%%%%%%%%%%%%%%%%%%%%%%%%%%%%%%%%%%%%%%%%%%%%%%%%%%%%%
\vspace{-0.1in}
\section{Introduction and Related Work}

%provide a clear statement of the problem and what the contribution of the work is to the relevant research community;
%state why this contribution is significant (what impact it will have);
%provide citation of the published literature most closely related to the manuscript; and
%state what is distinctive and new about the current manuscript relative to these previously published works.

%1Big picture Significance;
%2status quo;gap;
%3contribution;conclusions 
%4contributions.

\IEEEPARstart{S}{ounds}, also called acoustic concepts, are essential to how humans perceive and interact with the world. Sounds are captured in recordings, mainly on videos and the acoustic information captured is exploited in a number of applications. The dominant application is multimedia video content analysis, where audio is combined with images and text ~\cite{schauble2012multimedia,yebbn,lan2013cmu,cheng2012sri} to index, search and retrieve videos. Another application is Human-Computer interaction robotics~\cite{maxime2014sound,janvier2012sound} where humans and specially visually impaired and blind computer users~\cite{Edwards:1994:ATG:192426.192443} complement speech with sounds (e.g. laughing, clapping) as non-verbal communication. Recently, a growing application is in smart homes where sounds such as \textit{water tap running} are detected~\cite{Mesaros2016_EUSIPCO} and cities~\cite{Salamon:UrbanSound:ACMMM:14,yang2013psychoacoustical,hiramatsu2000response}, where sounds are investigated to detect acoustic pollution. All of these applications rely on automatic recognition of audio concepts to identify the occurrence of sounds within recordings.

Sound recognition research in the past few years has been furthered by competitions and standard datasets. From 2010 to 2015, the TRECVID-MED competition evaluated multimedia event detection on videos by analyzing sounds, images and text ~\cite{schauble2012multimedia,yebbn,lan2013cmu,cheng2012sri}. Moreover, the DCASE competitions in 2013~\cite{giannoulis2013detection} and 2016~\cite{Mesaros2016_EUSIPCO} evaluated scenes and sounds recognition in audio-only recordings. Additionally, the most popular standard datasets have allowed sound recognition to be tested in different contexts. For example, environmental sounds in the 2015's ESC-50~\cite{ESC50}, urban sounds in the 2014's US8K~\cite{UrbanSound} and more recently YouTube videos in the 2017's AudioSet~\cite{AudioSet} and DCASE 2017~\cite{DCASE2017challenge}. The approaches derived from research in these datasets have shown how well we can identify acoustic content corresponding to a label.

Labels in these datasets define sounds, but rarely describe nuances of sounds. Authors in~\cite{ntalampiras2017transfer,Stevenson2008} showed that audio is associated to a subjective meaning. For instance, in the DCASE dataset~\cite{giannoulis2013detection} there were two particular labels: \textit{quiet street} and \textit{busy street}. Sound recognition results and confusion matrices across papers~\cite{giannoulis2013detection,Mesaros2016_EUSIPCO} evidenced how although both labels defined audio from streets, the qualifier implied differences in the acoustic content. These kind of nuances can be described with different lexical combinations and two types that have been suggested in the literature are Adjective-Noun Pairs (ANPs) and Verb-Noun Pairs (VNPS).

Adjective-Noun Pairs can elicit a subjective meaning of the audio to the listener, defined by an adjective that shapes the natural and social environment~\cite{fruhholz2016sound}. Moreover, the subjectivity can cover other areas such as affective dimensions as explored by the authors of~\cite{Stevenson2008} where they collected the International Affective Digitized Sounds (IADS) dataset consisting of 111 sounds without enforcing a subjective word in the label. The sounds were presented to participants who had to categorize them into one of five classes: happiness, anger, sadness, fear, and disgust. Results showed how participants have consistent trends categorizing these sounds, suggesting a relationship between adjectives and audio content.

Verb-Noun Pairs can describe interactions between one or several objects and the material the objects are made of~\cite{darvishi1995designing,Schafer1993,Gygi2004}. For example, the interaction of objects and surfaces was explored in~\cite{owens2016visually}, where authors collected sounds corresponding to the action of drumsticks hitting and also scratching on different surfaces such as glass, wood and metal. Results suggested acoustic differences depending on the combination of the action, defined by a verb, and the surface, defined by a noun (e.g. scratching wood and scratching metal).

Investigation of adjectives and verbs has been successfully approached in other fields. In computer vision, Borth et al~\cite{Borth2013} introduced the VisualSentiBank to perform sentiment analysis of images~\cite{chen2014deepsentibank} based on Adjective Noun Pairs. In video analysis, actions described by verbs have been widely explored as described in these surveys~\cite{chaquet2013survey,poppe2010survey}. In text and language processing, authors in~\cite{baccianella2010sentiwordnet} introduced SentiWordnet to perform opinion mining using adjectives. In the music domain, acoustic characteristics and lyrics have been combined to detect sentiment in~\cite{zhong2012music}. Therefore, similar exploration on audio concepts will reveal to what extent we can automatically identify this information and how it could be combined for analysis of subjectivity in multimedia content.~\cite{picard1995computer,soleymani2015asm}.
%In~\cite{el2011survey}, authors presented a survey on speech emotion recognition and how different applications from HCI and Machine Translation benefit from this additional information.  such as \textit{good} or \textit{deplorable}~\cite{siersdorfer2010analyzing}

In this work we investigated for the first time the relation between audio content and both adjective-noun pair and verb-noun pair labels. The consistency between these type of pairs-based labels and audio can help to analyze sentiment, affect and opinion mining in audio as well as to complement similar pairs in other modalities such as images. Due to the lack of datasets with these types of annotations for audio, we collected, processed and released AudioPairBank. This is a large-scale corpus consists of 1,123 pairs-- 761 ANPs and 362 VNPs and over 33,000 audio files and its based on the collaborative repository called \textit{freesounds.org}. For this contribution we documented the challenges and implications of collecting audio recordings with these labels. These guidelines were not previously available in the literature and now serve as a direction for researchers to further annotate or refine our annotations. Another contribution is to show the degree of correlation between the audio content and the labels through sound recognition experiments and hence also providing a performance benchmark. Performance was better than random (less than 1\%) and around 70\% accuracy, considering the subjective nature of our labels. 

\vspace{-0.1in}

%Machine Hearing, as described by R. Lyon in~\cite{lyon2010machine}, aims to develop the AI capable of learning what is heard, name recognizable objects, actions, events and places as well as retrieving audio by reference to those names. The machines and systems with this AI should be able to listen and react in real time, to take appropriate actions and to naturally interact with humans. 

%%%%%%%%%%%%%%%%%%%%%%%%%%%%%%%%%%%%%%%%%%%%%%%%%%%%%%%%%%%%%%
%
%	Aqcusisition of the corpus
%
%
%
%%%%%%%%%%%%%%%%%%%%%%%%%%%%%%%%%%%%%%%%%%%%%%%%%%%%%%%%%%%%%%
\section{Collecting and Processing the AudioPairBank}
\label{collection}

\begin{figure}
\centering
\includegraphics[width=\linewidth]{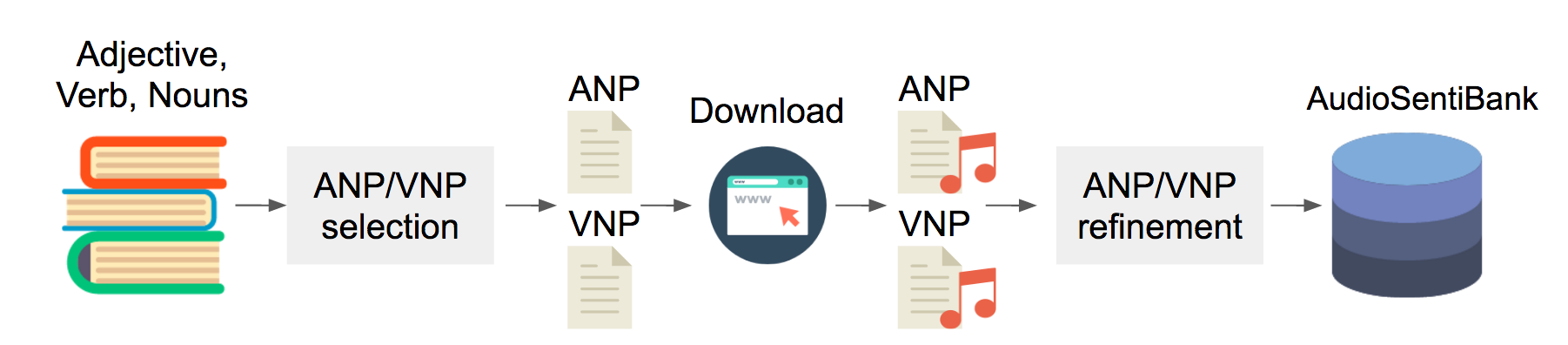}
\caption{An overview of the collection process of AudioPairBank. The ANPs and VNPs labels are based on existing ontologies. The labels are used as queries to download audio from \textit{freesound.org}. The labels and audio recordings were refined to create the final version of AudioSentibank.}
\label{fig:dataset_construction}
\end{figure}

We start by describing the steps for collecting and processing the corpus, which is illustrated in Figure \ref{fig:dataset_construction}. In Section~\ref{vocabulary}, we define the list of adjective-noun and verb-noun pairs based on existing ontologies. Then, in Section~\ref{download} we use these labels as queries to download audio recordings from an on line repository. Finally, we refine the dataset to reduce biases, outliers and implausible pairs in Section\ref{filter} to output the finalized AudioPairBank. The detailed process of each step can be seen in~\cite{sebastianAudio}.

\subsection{Selecting ANPs and VNPs based on ontologies}
\label{vocabulary}

We looked into existing ontologies containing adjectives, nouns and verbs to collect a list of 10,829 pairs for adjective-nouns and 9,996 pairs for verb-nouns. An ontology by Davies~\cite{Davies2013} defined three audio semantic levels: sound sources, sound modifiers and soundscape modifiers based on a research where participants were asked to describe sounds based on nouns, verbs and adjectives. Additionally, Axelsson~\cite{Axelsson2010} suggested a list of adjectives to describe the feeling that sounds produce to individuals. Another pair of ontologies introduced by Schafer in~\cite{Schafer1993} and Gygi~\cite{Gygi2004} are based on soundscapes and environmental sounds, where sounds were labeled by their generating source using verbs, such as \textit{Baby crying} and \textit{Cat meowing}. Lastly, we considered the Visual Sentiment Ontology (VSO) presented in~\cite{Borth2013}, which is a collection of ANPs based on 24 emotions defined in Plutchik’s Wheel of Emotions.

After an inspection of the final list of pairs we noticed lexical variations that were grouped. One example are comparatives and superlatives such as \textit{faster} and \textit{fastest}, which were grouped together as \textit{fast}. Another example are 
synonyms such as \textit{car}, \textit{auto}, \textit{automobile}, which were grouped together as \textit{car}. Another example are plural forms such as \textit{dogs} or \textit{cars}, which were grouped into their singular forms. This process implied assumptions in the audio content which may not hold true. For instance, the sound of one car is acoustically different from the sound of multiple cars. Nevertheless, grouping helped to reduce the impact of having multiple ``repeated'' pairs which could result in major acoustic ambiguities and low sound recognition performance. %Later on we found that even if we would not have grouped these words together, the co-occurrences of these tags would have resulted in retrieving nearly the same audio files.

\subsection{Downloading ANPs and VNPs from the web}
\label{download}

The list of adjective-noun and verb-noun pairs from the previous section was used to query and download audio recordings from \textit{freesound.org}. The website has the largest audio-only archive of sounds with around 230,000 recordings, which allowed us to collect audio in a large-scale. Moreover, the website has been successfully employed before to create popular datasets (ESC-50, Freefield,UrbanSounds)~\cite{ESC50,Freefield1010,UrbanSound}. Other websites such as \textit{soundcloud.com} and \textit{findsounds.com} were considered, but not employed because they either contained mainly music or had less sound recordings in their archives.

The \textit{freesound.org} website is a collaborative repository of audio where users upload recordings and write tags to describe their content. This folksonomy-like structure of repository has the benefit of reflecting the popular and long-tailed combinations of tags. In this manner we can observe what are the socially-relevant adjectives, verbs and nouns. 

The tags of a given recording are combined to create weak labels of adjective-noun pairs and verb-noun pairs. The weak labels happen because users upload audio recordings and provide tags based on what they consider relevant. However, tags do not follow a particular order, may describe only a portion of the content and are not accompanied with the time-stamp of the sound-tag occurrence. Moreover, the order of the tags could change the meaning, such as a verb could be intended to rather be an adjective. The intent of the user is unknown, but its exploration is necessary for large-scale collection and analysis. Also, we expect the machine learning algorithms to help us determine the degree of relation between such pairs and their corresponding sounds.

%%%%%%%%%%%%%%%%%%%%%%%%%%%%%%%%%%%%%%%%%%%%%%%%%%%%%%%%%%%%%%
%
%	Filtering steps performed to create the final corpus
%
%
%
%%%%%%%%%%%%%%%%%%%%%%%%%%%%%%%%%%%%%%%%%%%%%%%%%%%%%%%%%%%%%%
\subsection{Refining the downloaded ANPs and VNPs}
\label{filter}

%\begin{figure}[t]
%\centering
%\includegraphics[width=0.45\linewidth]{images/sample_rate_distribution_ANPs}
%\caption{Distribution of the sampling rates of the ANP data.}
%\label{fig:sampling_rates}
%\end{figure}

\begin{figure}[t]
\centering
\includegraphics[width=\linewidth]{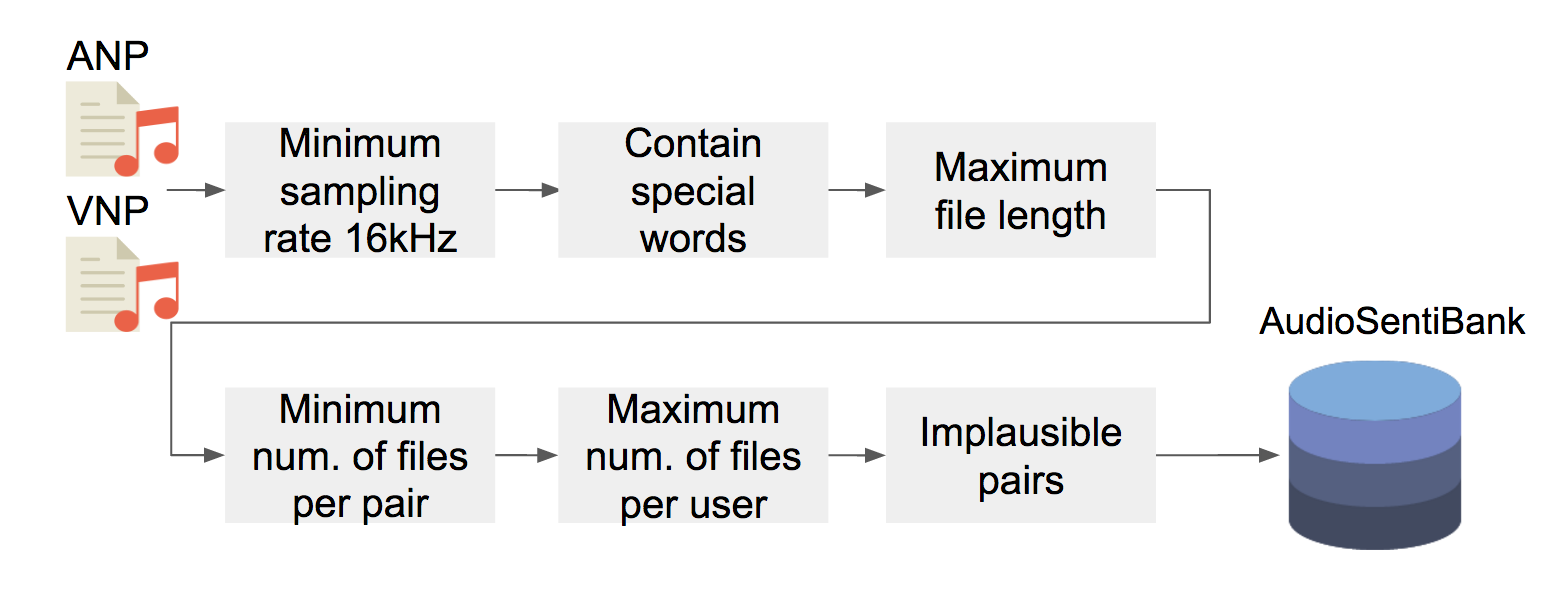}
\caption{The refining process consisted of six steps where pairs that did not meet a certain criteria were filtered out. Steps 2,5 and 6 were responsible of the highest number of discarded pairs.}
\label{fig:data_filtering}
\end{figure}

The downloaded audio recordings along with their labels revealed several characteristics discussed in this section. Therefore, we refined the corpus with the goal of increasing the quality and diversity of audio concept pairs. The process is illustrated in Figure~\ref{fig:data_filtering}. A manual revision has been employed by other authors~\cite{Borth2013,AudioSet} to improve their automatically collected datasets. 

\paragraph{Minimum sampling rate} The chosen rate was 16 kHz and files with a lower rate were discarded. The value was selected because lower rates restrict the amount of frequency information and because 95\% of the files were above 16kHz and mainly 44.1kHz. 

\paragraph{Contain special words} We removed pairs tagged with words that implied unwanted content. For example, \textit{loop}, \textit{loops} or \textit{looping} contained sounds which repeated over and over again. The repetition or periodicity was artificial and could mislead the sound recognition systems. Another example were pairs with words such as \textit{sound}, \textit{audio} or \textit{effect} because they did not add meaning. A third example were pairs with the word \textit{processed} because the audio files commonly contained music or overlapped music throughout the recording. We also removed pairs with redundancy of terms such as \textit{noisy noise} or \textit{natural nature}. Another discarded pattern happened with terms related to music genre such as \textit{heavy metal} and \textit{classic rap} or types such as \textit{waving techno} and \textit{ringing music}. Nevertheless, we kept some pairs related to music such as \textit{happy music}, \textit{sad music} or \textit{dramatic guitar}. Another pattern happened with sound packs, which consisted of groups of audio files, with exactly the same tags and uploaded by the same user. However, not all audio content from every file was related to the tags. Because these bundles did not occur often and are hard to track automatically, we removed the obvious ones, but perhaps kept others.

\paragraph{Maximum audio file length} We removed recordings with long duration because longer audio files had audio content that was not described by the tags. We computed the distribution of the duration for each pair and removed the outliers in the distribution based on Tukey's range test\footnote{\url{http://en.wikipedia.org/wiki/Tukey\%27s_range_test}}. We defined outliers as values larger than the third quartile of the distribution plus 1.5 times the Interquartile Range (IQR), and formally: $outlier > Q3 + 1.5 * IQR$. The average length was around 20 seconds, but pairs related to field-recordings had some files longer than 900 seconds (15 minutes), which were also discarded.

\paragraph{Minimum number of files per pair} We discarded pairs with less than 20 files. because such low count was an indicator of rare pairs, which might not be worth exploring in this work. The minimum number is consistent with other available datasets~\cite{piczak2015environmental,Mesaros2016_EUSIPCO}.

\paragraph{Maximum number of audio files per user} For some pairs there were users who dominated the contribution of audio recordings. Users tend to use similar recording devices and conditions causing a data bias. As a consequence, machine learning-based algorithms for sound recognition can learn these biases instead of the audio content as demonstrated in~\cite{lei2011user}. Therefore, we reduced the influence of a user by allowing a maximum of 25\% of recordings per pair. 

\paragraph{Implausible pairs: manual inspection and plausibility score}
\begin{comment}
%For the ANPs there were several common terms such as \textit{slow motion} or \textit{fast food} which usually have a different meaning than our ANP because the ANP is made up of two tags that a user added individually to the recording.
But we removed following ANPs/VNPs because some of them are terms related to music and some are just implausible:
ANPs
heavy metal
extreme metal
classic music
weird techno
industrial techno
extreme techno
echoing techno
heavy techno
strange techno
loud techno
fast techno
classic techno
classic rap
slow hiphop
classic hiphop
industrial music

VNPs
clapping techno
waving techno
breaking techno
breathing techno
clapping music
talking music
ringing music
walking music
\end{comment}

\begin{figure}
\centering
\includegraphics[scale=0.42]{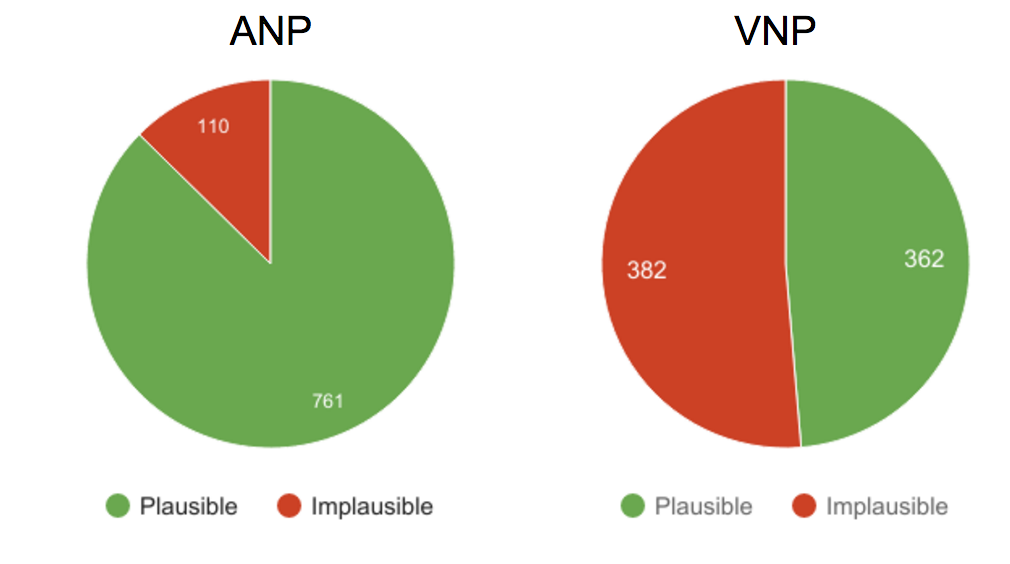}
\caption{The charts show the number of plausible and implausible pairs after using both methods, manual inspection and plausibility score. In contrast to adjectives, verbs have more restrictions in how they can be combined with nouns to convey plausible meaning.}
\label{tab:plausible_statistics}
\end{figure}

Manual inspection was used to catch salient implausible pairs and was complemented with a data-driven metric to determine the degree of plausibility. Both approaches were responsible of discarding 12\% of the ANPs 51\% of VNPs as shown in Figure~\ref{tab:plausible_statistics} and some examples are in Table~\ref{tab:plausible_examples}.

We manually inspected the audio concept pairs and discarded those which were implausible and could not be associated to an acoustic semantic meaning. For example, some ANPs derived from Plutchik’s Wheel of emotion, such as \textit{slow fear}, a sound that is arguably impossible to reproduce. This problem was also faced in the Visual Sentiment Ontology (VSO)~\cite{Borth2013} and pairs were discarded, such as \textit{fresh food} or \textit{favorite book}. Other examples of implausible pairs are \textit{walking winter} and \textit{singing park}, where nouns define a time and place, but turned implausible because of the verb. Some implausible nouns like \textit{future} or \textit{design} represented abstract meanings and were hardly connected to consistent sounds. An interesting pair also discarded was \textit{talking bird}, which appeared to be semantically wrong, but possible. A closer look into the corresponding audio files revealed recordings of a bird talking, a parrot perhaps, but these recordings were rare and the majority of the recordings contained talking people with bird sounds in the background.

%Additionally, we observed an overlap of concept pairs as for example we had \textit{talking man} or \textit{talking child} and on the other side there is \textit{talking english} or \textit{talking spanish}. Hence, the same recording could be tagged with multiple concept pairs and make the classification ambiguous. In this case we decided to keep them, because the acoustics of from children's speech and man's speech are sufficiently different, as well as the phonetics from languages.

The plausibility score was designed to favor diversity of users, number of files and uniqueness of files for the given pair. For example, the manually discarded pair \textit{singing park} was selected because a park cannot sing. This pair might have emerged because files were tagged with different words such as \textit{singing}, \textit{walking}, \textit{relaxing}, \textit{park}, \textit{bird} and \textit{people}. Although the pair \textit{singing park} occurred in several audio files, it never occurred together in any other recording and was rather always together with other pairs such as \textit{singing bird} and \textit{walking people} and \textit{relaxing park}. Hence, these kind of pairs yielded a low plausibility score. Formally the score is defined as following:

\begin{equation}
PS(cp) = \frac{\frac{u_{cp}}{n_{cp}}+\frac{f_{cp}}{n_{cp}}}{2} = \frac{u_{cp}+f_{cp}}{2n_{cp}}
\label{eq:plausibility_score}
\end{equation}

\vspace{-0.1in}

where $n_{cp}$ is the total number of files belonging to a concept pair $cp$. Then, $u_{cp}$ is the number of unique users that uploaded files for a concept pair $cp$ and $f_{cp}$ is the number of files that are unique to a concept pair $cp$. A file is unique to a concept pair if it is not tagged with any other concept pair in the existing set. The division by $2$ is necessary because for a perfectly plausible concept pair the numerator becomes $1+1$ and so the division by two will keep the metric in the range between $0$ (least plausible) and $1$ (most plausible). We observed that rare pairs such as the ones described in the previous paragraphs obtained a score lower than 0.2. 

\begin{table}[t]
\centering
\caption{Examples of plausible and implausible pairs. Note how for example ``fast food'' is an implausible pair because despite it could be associated to a semantic meaning it does not convey an acoustic semantic meaning.}
\begin{tabular}{c|c|c|c} 

\textbf{Plausible} & \textbf{Implausible} & \textbf{Plausible} & \textbf{Implausible} \\ 
\textbf{ ANPs} & \textbf{ ANPs} & \textbf{ VNPs} & \textbf{ VNPs} \\
\hline happy music & windy bird & singing bird & laughing animation \\ 
 slow car & industrial hands & crying baby & falling autumn \\ 
 echoing footsteps & slow fear & flying bee & clapping hat \\ 
 rattling machine & fast food & honking car & talking text \\  
\end{tabular}
\label{tab:plausible_examples}
\end{table}

\subsection{Finalized AudioPairBank}
\label{final}

The refined corpus is one of the largest available datasets for sounds and the only one with adjective-noun and verb-noun labels. The ANPs and VNPs are weak labels based on the collaborative repository in \textit{freesound.org}. The main statistics of the corpus are included in Table \ref{tab:filtered_statistics}. AudioPairBank consists of 761 ANPs and 362 VNPs for a total of 1,123 pairs. One or more pairs can correspond to the same audio file. The number of shared-unique audio files are 58,626-16,335 for ANPs and 38,174-20,279 for VNPs for a total of 96,800-33,241 files. The average number of unique file are 21 for ANPs and 56 for VNPs. The number of unique nouns is 1187, unique verbs is 39 and unique adjectives is 75. The influence of a user per pair was set to a maximum of 25\% contribution of files. Note that the total number of unique files and users does not correspond to the sum of the previous rows because some files are repeated in both categories. Regarding the length, ANPs had a larger file duration than VNPs with almost twice the length. The last column shows the total disk space of the audio files in WAV format.
% BME Sebastian, number of files shared between anps and vnps , table of hours and size is for unique or total?

\begin{comment}
\begin{table}
\centering
\caption{Statistics of Audio SentiBank. The ANPs have more pairs, a higher total of files and a larger average duration per file. VNPs have more unique files, and more diversity of user contribution.}
\begin{tabular}{|c|c|c|c|c|c|}
\hline  & \textbf{Pairs} & \textbf{Total - Unique Files} & \textbf{Users} & \textbf{Length} & \textbf{Size} \\ 
\hline ANPs & 1,016 & 81,771 - 20,233 & 2,806 & 892 h & 1.4 TB \\ 
\hline VNPs & 824 & 72,424 - 26,347 & 3,905 & 375 h & 0.8 TB \\ 
\hline Total & 1,840 & 154,195 - 41,993 & 5,080 & 1,267 h & 2.2 TB \\ 
\hline 
\end{tabular}
\label{tab:complete_statistics}
\end{table}
\end{comment}

%while the statistics of the complete pre-refined dataset are given in Table \ref{tab:complete_statistics}.

\begin{table}[t]
\centering
\caption{ANPs double the number of VNPs because audio recordings are generally tagged with more adjectives than verbs.}
\vspace{-0.2cm}
\begin{tabular}{c|c|c|c|c|c}
& \textbf{Pairs} & \textbf{Total - Unique Files} & \textbf{Users} & \textbf{Hours} & \textbf{Size} \\ 
\hline ANPs & 761 & 58,626 - 16,335 & 2,540 & 892 & 528 GB \\ 
 VNPs & 362 & 38,174 - 20,279 & 3,279 & 375 & 212 GB \\ 
 Total & 1,123 & 96,800 - 33,241 & 4,478 & 1,267 & 740 GB \\  
\end{tabular} 
\label{tab:filtered_statistics}
\end{table}

%Since we filtered the data by several constraints, we compared the filtered data to the final data with respect to several metrics. Therefore, Table \ref{tab:statistics_duration_tags} shows how the mean and median of the duration and tags per file changed. As expected, the duration decreased, but the effect is much stronger for the mean than for the median because we only removed extreme values and the median was robust. The number of tags per file decreased, but only slightly for mean and median.

%Next step was to contruct an ontology defining a taxonomy for the nouns and discovering relationships between adjective and verbs and their interplay with nouns. This ontological structure will be of great use to define scenarios using AudioPairBank.

\begin{comment}
\begin{table}[t]
\centering
\caption{Statistics of durations and tags}
\begin{tabular}{|c|c|c||c|c|}
\hline
\multicolumn{1}{|c|}{} & \multicolumn{2}{|c||}{\textbf{Duration per File}} & \multicolumn{2}{|c|}{\textbf{Tags per File}} \\
\hline  & \textbf{Mean} & \textbf{Median} & \textbf{Mean} & \textbf{Median} \\ 
\hline ANPs (complete) & 112.84 & 14.77 & 21.84 & 19 \\ 
\hline ANPs (filtered) & 107.8 & 14.75 & 21.21 & 18 \\ 
\hline VNPs (complete) & 70.88 & 8.91 & 15.48 & 12 \\ 
\hline VNPs (filtered) & 58.72 & 7.87 & 14.59 & 11 \\ 
\hline 
\end{tabular} 
\label{tab:statistics_duration_tags}
\end{table}
\end{comment}

\vspace{-0.1in}

%%%%%%%%%%%%%%%%%%%%%%%%%%%%%%%%%%%%%%%%%%%%%%%%%%%%%%%%%%%%%%
%
%	Analysis of the corpus
%
%
%
%%%%%%%%%%%%%%%%%%%%%%%%%%%%%%%%%%%%%%%%%%%%%%%%%%%%%%%%%%%%%%
\section{Analysis of AudioPairBank}
\label{statistics}

We performed an analysis on different aspects of ANPs and VNPs, such as co-occurrences of adjective, verb, nouns, also duration of audio recordings, number of audio files, number of tags, and number of users. 

\subsection{Number of audio files per ANP and VNP}
\begin{figure*}[t]
\centering
\subfloat[ANPs with at least 400 files]{\includegraphics[width=.44\textwidth, height=6cm]{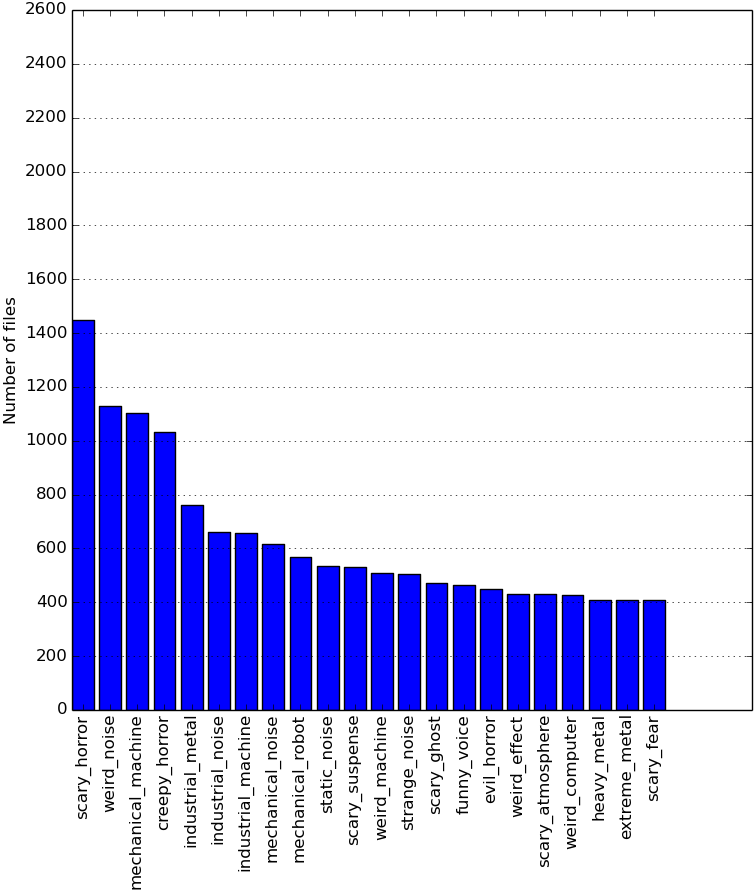}}~~~~
\subfloat[VNPs with at least 400 files]{\includegraphics[width=.44\textwidth, height=6cm]{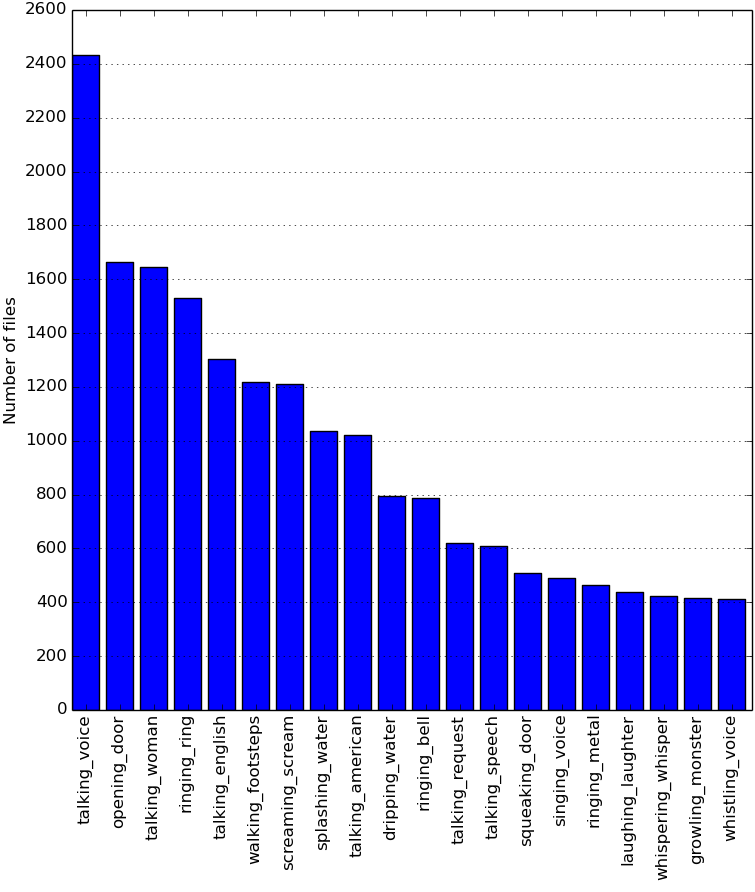}}
\caption{Both plots show a decreasing distribution with a long tail trimmed due to space limitations. ANPs show a smoother decrement, which translates to a more uniform number of files per concept pair.}  
\label{fig:concept_pair_distribution}
\end{figure*}

The distribution of the number files per pair for ANPs and VNPs shows that some pairs are more common than others. This information is useful because it gives us an intuition of which pairs are more common for users on the web. Figure~\ref{fig:concept_pair_distribution} shows a decreasing distribution with a long tail trimmed due to space limitations. ANPs show a smoother decrement, which translates to a more uniform number of files per concept pair.

\begin{figure*} [htb]
\subfloat[ANPs with at least 400 files]{\includegraphics[width=.48\textwidth, height=6cm]{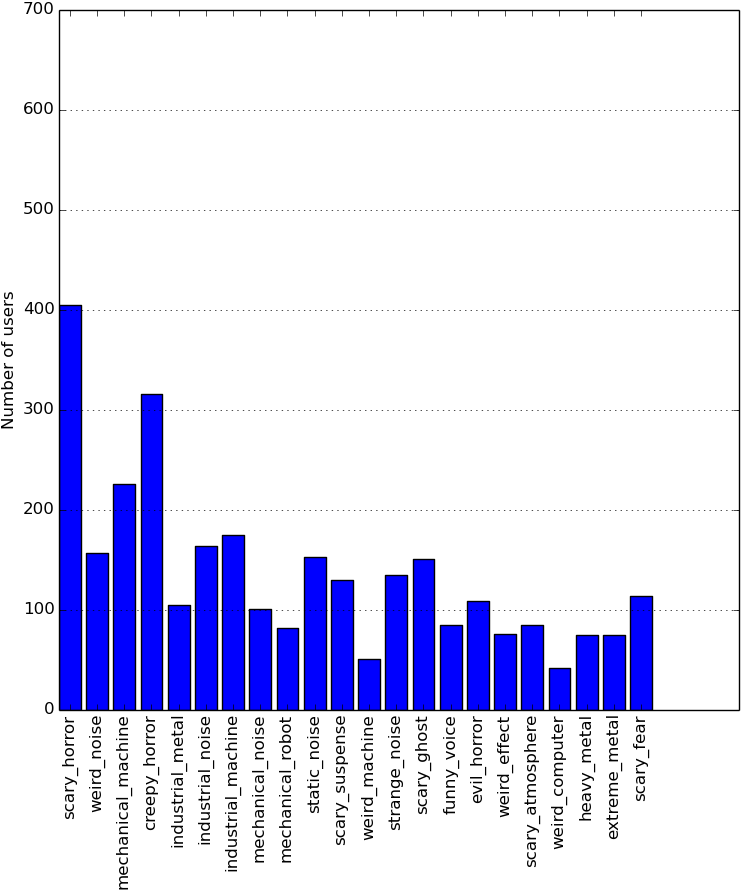}}~~~
\subfloat[VNPs with at least 400 files]{\includegraphics[width=.48\textwidth, height=6cm]{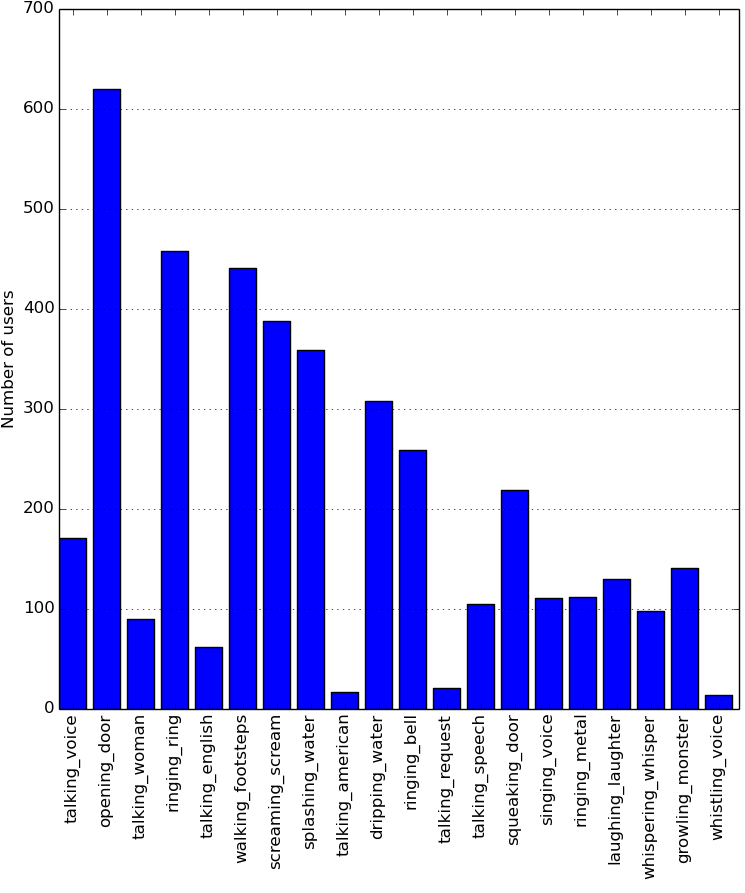}}
\caption{The number of users that uploaded files corresponding to a concept pair are color-coded, showing the diversity of users per concept pair.}  
\label{fig:concept_pair_user_distribution}
\end{figure*} 

\subsection{Number of users per ANP and VNP}
\begin{figure*}[t]
\subfloat[ANPs with at least 400 files]{\includegraphics[width=.48\textwidth, height=6cm]{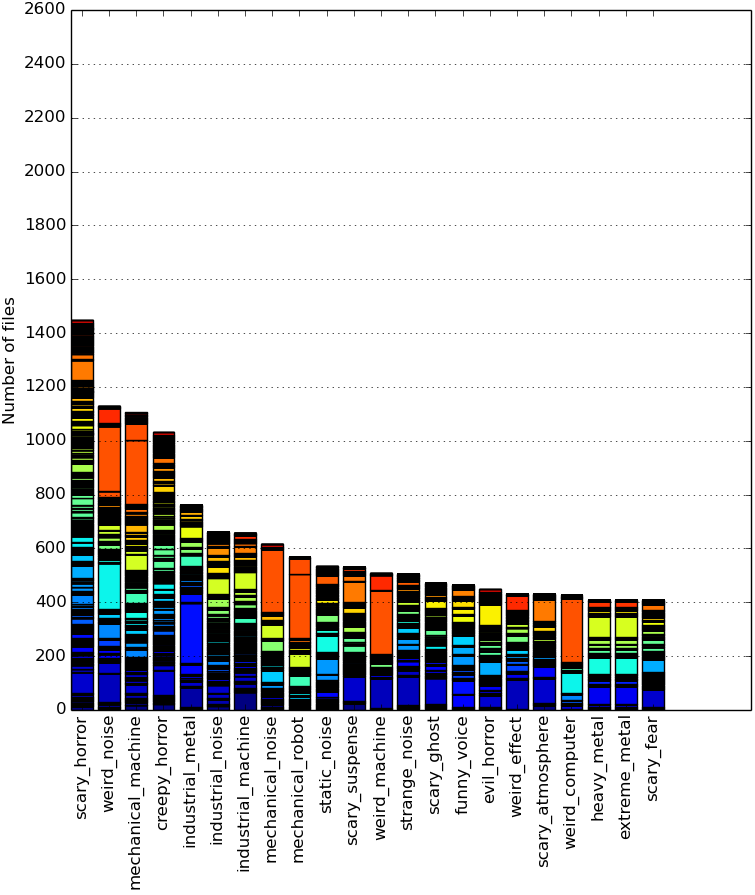}}~~~~
\subfloat[VNPs with at least 400 files]{\includegraphics[width=.48\textwidth, height=6cm]{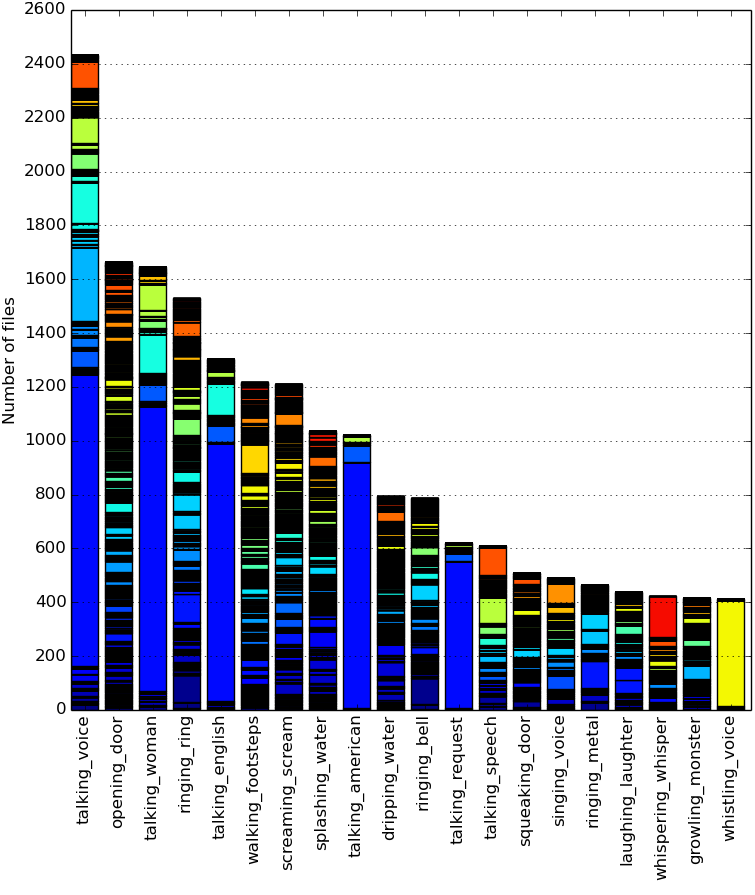}}
\caption{Number of files per concept pair color-coded by the uploading users. The Figure shows diversity of user's contribution per concept pair despite some exceptions.}  
\label{fig:concept_pair_usercoded_distribution}
\end{figure*}

We looked at how many different users contributed to the pairs, which is illustrated in Figure~\ref{fig:concept_pair_user_distribution}. This gives an intuition about the diversity of users per concept pair. As expected, some pairs had more contributors than others. Moreover, Figure~\ref{fig:concept_pair_usercoded_distribution} helps to visualize how the files per pair are distributed among users.

We also observed that most users commonly employed a small set of tags with high frequency. The most frequent tags across users are shown in Table~\ref{tab:frequent_tags}. Rarely occurring tags often come from single users.

\begin{table}[h]
\centering
\caption{High-frequency tags common across users.}
\begin{tabular}{c|c}
      & \textbf{Tags}  \\ 
\hline Adjectives & \textit{scary}, \textit{creepy},   \textit{industrial}, \textit{funny} \\ 
Verbs & \textit{talking}, \textit{walking}, \textit{laughing}, \textit{singing},   \\  Nouns & \textit{atmosphere}, \textit{horror}, \textit{noise}, \textit{voice} \\ 

\end{tabular} 
\label{tab:frequent_tags}
\end{table}

\begin{comment}
in the ANP data: 20 tags 761 = 15,220
    funny 2767
    scary 7949
    creepy 5501
    industrial 3760
in the VNP data: 15 tags 362 5,430
    laughing 1237
    singing 1259
    talking 7271
    walking 1857
\end{comment}

\subsection{Duration of ANPs and VNPs audio files}
%include for AVNs: avg size, length variance.
Adjectives, in some cases, hint the duration of the recording. For example on average, audio containing \textit{calm}, \textit{rural}, \textit{peaceful} and \textit{quiet} had longer durations (more than five minutes) than those tagged with \textit{accelerating} or \textit{rushing} (less than two minutes).

Verbs describe actions and hence we expected the duration of the audio recordings of VNPs to correspond to the approximate length of the described action. However, even if most actions lasted between one to five seconds, the median duration of VNP audio files was around 10 seconds. This may happen because the action described by the verb tend to occur more than once, or because other actions take place within the recording.

Nouns have more length variability that depends on what it is describing. Locations, environments and field recordings are associated to longer durations, such as \textit{city}, \textit{market}, \textit{beach} or \textit{rain}, while objects are associated to shorter durations, such as \textit{cup} or \textit{door}.

\subsection{Correlation between the number of tags and the length of the ANPs and VNPs audio files}
\begin{figure*}[t]
\subfloat[Number of tags vs duration of files corresponding to ANPs]{\includegraphics[width=1\textwidth, height=6cm]{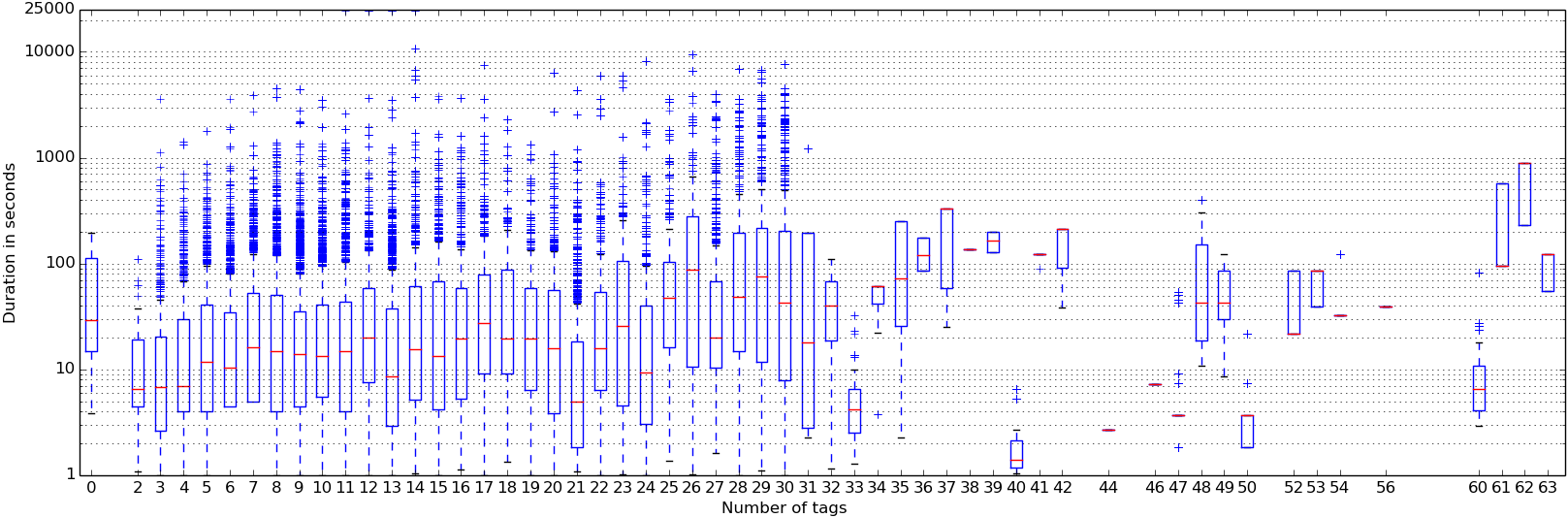}}~~~~
\caption{The audio file duration and the number of tags showed almost no correlation between them based on the Spearman’s rank correlation coefficient. The average number of files have between 10 to 20 tags. A similar trend was observed for VNPs.}  
\label{fig:concept_pair_ntags_durations}
\end{figure*}

We looked at the correlation between the number of tags and the length of the file corresponding to a pair, because we expected that more tags will translate into longer duration. However, we found a weak to almost no correlation between both, as illustrated in Figure \ref{fig:concept_pair_ntags_durations}. We validated our observations by computing Spearman’s rank correlation coefficient (SRCC), which is a non-parametric measure of statistical dependence between two variables. An SRCC value close to zero in combination with a very small p-value indicates no correlation in the data. In our case, the SRCC value for ANPs is 0.073 with a p-value of 5.358e-97, and for VNPs is 0.0149 with a p-value of 6.31e-05. Nevertheless, the number of tags together with the duration of the audio file suggested a distinction between acoustic scenes/soundscapes and sounds.

\subsection{Co-occurrences of tags in ANPs and VNPs audio files} 
In order to understand the context in which concept pairs occur, we analyzed the co-occurrences of the accompanying adjectives, nouns and verbs tags within the audio file. %A reason might be because users used multiple similar terms to attempt to have a broader description of their files and also to increase the keyword-matching hits. 

Some adjectives occur more frequently than others, such as \textit{loud}, \textit{heavy}, \textit{scary} and \textit{noisy} in contrast to \textit{exotic}. Adjectives that occur frequently tend to describe nouns that are commonly locations: \textit{landscape}, \textit{coast} or \textit{nature}. As expected, we found almost no tags using colors as adjectives. An interesting co-occurrence of adjectives was when they had opposite meaning such as \textit{slow} and \textit{fast} or \textit{peaceful} and \textit{loud}. After manual inspection of audio files we concluded that this was an indicator of changes in the audio content throughout the recording, specially when the noun was shared. For example, a recording had two pairs, \textit{slow train} and \textit{fast train}, which had a train passing slowly and then followed by another one passing at high-speed.

Verbs tend to occur less frequently than adjectives and with a more restricted set of nouns, for instance, \textit{flying} occurs mainly with \textit{airplane}, \textit{engine}, \textit{bird}, and \textit{helicopter}. In a similar manner, almost the same verbs occur with human-related nouns such as \textit{baby}, \textit{child}, \textit{man} and \textit{woman}. One pattern observed for verbs is that they may describe and complement another verb. For example, \textit{open} and \textit{close} co-occur with \textit{squeaking} and \textit{banging}, all with the noun door. These combinations specify how the door was opened or closed. Similar to adjectives-adjectives, verbs could be an indicator of changes in the audio content throughout the recording, such as \textit{singing} and \textit{clapping} described a music concert. %Finally, another pattern happened when verbs served to clarify other verbs. 

Nouns were more helpful to provide context and clarify the sound source. An example of context are \textit{thunder}, \textit{rain} and \textit{wind}, which described the acoustics of a storm, also \textit{frogs} and \textit{insects} and \textit{water} described the acoustics of a Savannah. Moreover, users sometimes included the location such as \textit{Florida} or the time such as \textit{day} and \textit{night} or the season such as \textit{spring}. An example of sound source is when the sound of an \textit{engine} happened with \textit{car} or \textit{train} or \textit{airplane} or \textit{ship}, and thus we knew the specific source of the engine sound. It is important to mention the noun \textit{noise}, which co-occurs very frequently with other tags, sometimes indicates an unintelligible sound, but was more commonly employed to define sounds happening in the background which were unrelated to the target sound.

\begin{comment}
\begin{figure*}[t]
\subfloat[[VNP Tag Distribution]{\includegraphics[width=1\textwidth, height=6cm]{images/tag-count_duration_VNPs}}
\caption{Distribution of audio duration for files with different number of tags. The audio file duration and the number of tags showed no correlation between them, as supported by Spearman’s rank correlation coefficient.}  
\label{fig:concept_pair_ntags_durations}
\end{figure*}
\end{comment}

\vspace{-0.1in}

%%%%%%%%%%%%%%%%%%%%%%%%%%%%%%%%%%%%%%%%%%%%%%%%%%%%%%%%%%%%%%
%
%	Experiments & Benchmarksing 
%
%
%
%%%%%%%%%%%%%%%%%%%%%%%%%%%%%%%%%%%%%%%%%%%%%%%%%%%%%%%%%%%%%%
\section{Experimental Setup for Benchmark Computation of AudioPairBank}
\label{experiments}

The previous sections described the challenges of collecting acoustic pairs, such as weak labeling from folksonomies. However, we also refined our pairs expecting to find consistency between the acoustics and the labels describing the actions (verbs) and properties (adjectives) of the sounds. Therefore, in this section we describe the setup of the two main experiments on the audio of the Adjective-Noun Pairs and Verb-Noun Pairs contained in AudioSentibank. First, the input audio was standardized into the same format and passed through a feature extraction step. Next, the extracted features were passed through a system for detection (or binary classification) and multi-class classification.

The audio files from the dataset needed to be standardized into the same format. We chose WAV format, sampling rate of 44.1 kHz, encoding PCM 16 bits and one channel. These parameters were already dominant in the corpus and are also common in the available datasets. Moreover, audio files have variable length and thus, each file was trimmed into four second segments with a 50\% overlap. These two parameters yielded the best sound event detection results among different values~\cite{Chu2009,UrbanSound}. The AudioSentibank corpus has three partitions, Training, Cross Validation (CV) and Testing with a ratio of 40\%-30\%-30\%. 

For each audio, we extracted Mel Frequency Cepstral Coefficients (MFCCs) features because they provide a competitive baseline performance. We used the toolbox Yaafe~\cite{Yaafe} to compute MFCCs with 13 and 20 coefficients and appended their first and second derivatives to include the dynamic information. The window size was 25 ms every 10ms. %BME Check last sentence with Sebastian

Detection allows the classifier to decide whether a pair-class is present or not. We used a one-vs-all setup using a Support Vector Machine (SVM)~\cite{scikit-learn} classifier. We trained one SVM per each AN and VN pair using 100 segments corresponding to the positive (target) pair-class and 200 corresponding to a negative (not-the-target) pair-class. The 100 segments were randomly selected from the pool of the positive pair-class. This number corresponds to the minimum number of segments per pair-class. The 200 segments were randomly selected from the pool of other classes avoiding repeated files. The SVM had a linear kernel and using the CV set we tuned the soft margin parameter $C$ with the following values: 5.0, 2.0, 1.0, 0.5 and 0.01, with 5 yielding the best results. %BME Check last sentence with Sebastian

Multi-class classification has to discern between multiple pair-classes and forces every input audio to belong to one of the trained pair-classes and we employed two different algorithms. First, we used a multi-class Random Forest (RF)~\cite{scikit-learn} classifier used to compute baseline performance in~\cite{piczak2015environmental,Salamon:UrbanSound:ACMMM:14}. RFs are an ensemble learning method that operates by creating multiple decision trees at training time. Then, at testing time each tree votes to predict a class minimizing over fitting. We trained two RFs, one for all the VNPs and one for all the ANPs using 100 segments corresponding to each class. The 100 segments were randomly selected from the pool of positive segments. The two RFs were tuned using the CV set to find the number of trees, we tried 5, 10, 20, 50 and 100, where 100 yielded the best results. Using more than 100 trees surpassed the capabilities of our computing resources. Second, we trained a Convolutional Neural Network (CNN), which is the foundation of state of the art sound classification. In~\cite{piczak2015environmental}, the log-mel spectrogram of each sound event recording is treated as an image and passed to the CNNs. We employed the same architecture. The first convolutional ReLU layer consisted of 80 filters of rectangular shape (57x6 size, 1x1 stride) allowing for slight frequency invariance. Max pooling was applied with a pool shape of 4x3 an stride of 1x3. A second convolutional ReLU layer consisted of 80 filters (1x3 size, 1x1 stride) with max pooling (1x3 pool size, 1x3 pool slide). Further processing was applied through two fully connected hidden layers of 5000 neurons with ReLU non-linearity. The final output is a softmax layer. Training was performed using Keras implementation of mini-batch stochastic gradient descent with shuffled sequential batches (batch size 1000) and a nestrov momentum of 0.9. We used L2 weight decay of 0.001 for each layer and dropout probability of 0.5 for all layers.  

To evaluate detection performance we computed accuracy, f-score and AUC and for multi-class classification we computed accuracy.

\begin{comment}
\begin{table}[t]
  \centering
  \caption{Detection results with linear SVMs. More coefficients yielded better results.}
  \begin{tabular}{|c|c|c|c|c|}
    \hline
     & \textbf{Features} & \textbf{Accuracy} & \textbf{f-score} & \textbf{AUC} \\
    \hline
    \multirow{4}{*}{ANP} & 13 MFCCs+$\Delta$+$\Delta\Delta$ & 0.68 & 0.51 & 0.69 \\
	& 20 MFCCs+$\Delta$+$\Delta\Delta$ & 0.69 & 0.51 & 0.70 \\
	& 30 MFCCs+$\Delta$+$\Delta\Delta$ & 0.70 & 0.50 & 0.70 \\
	\hline
    \multirow{4}{*}{VNP} & 13 MFCCs+$\Delta$+$\Delta\Delta$ & 0.69 & 0.53 & 0.71 \\
   	& 20 MFCCs+$\Delta$+$\Delta\Delta$ & 0.71 & 0.53 & 0.72 \\
   	& 30 MFCCs+$\Delta$+$\Delta\Delta$ & 0.71 & 0.51 & 0.73 \\
    \hline
  \end{tabular}

  \label{table:detection_results}
\end{table}

\begin{table}[t]
  \centering
  \caption{Multi-class classification results with Random Forests. Contrary to detection results, less coefficients yielded better results.}
\begin{tabular}{|c|c|c|}
    \hline
     & \textbf{Features} & \textbf{Accuracy\%} \\
    \hline
	\multirow{2}{*}{ANP} & 13 MFCCs+$\Delta$+$\Delta\Delta$ & 1.6\\
     & 20 MFCCs+$\Delta$+$\Delta\Delta$ & 1.5\\
   	\hline
    \multirow{2}{*}{VNP} & 13 MFCCs+$\Delta$+$\Delta\Delta$ & 4.5 \\
     & 20 MFCCs+$\Delta$+$\Delta\Delta$ & 4.4 \\
    \hline
  \end{tabular}
  \label{table:multi_class_classification_results}
\end{table}
\end{comment}

\begin{table}[t]
  \centering
  \caption{Overall detection performance.}
  \begin{tabular}{c|c|c|c|c}
     & \textbf{Features} & \textbf{Accuracy\%} & \textbf{f-score\%} & \textbf{AUC\%} \\
    \hline
    \multirow{1}{*}{ANP} & 13 MFCCs+$\Delta$+$\Delta\Delta$ & 69 & 51 & 70 \\	
    \multirow{1}{*}{VNP} & 13 MFCCs+$\Delta$+$\Delta\Delta$ & 71 & 53 & 72 \\

  \end{tabular}

  \label{table:detection_results}
\end{table}

\begin{table}
  \centering
  \caption{The top 5 best and worst detected ANPs and VNPs.}
  \begin{tabular}{c|c|c|c}
    \multicolumn{2}{c|}{\textbf{ANP}} & \multicolumn{2}{|c}{\textbf{VNP}} \\
    %\hline
   	\textbf{Best} & \textbf{Worst} & \textbf{Best} & \textbf{Worst} \\
    \hline
    weird cup & funny english & howling dog & splashing water \\
    industrial phone & heavy rain & howling wolf & crackling footsteps \\
    echoing phone & noisy glitch & crying insects & splashing river \\
    echoing alert & extreme noise & howling animal & gurgling water \\
    weird cell & loud fireworks & ringing cup & breaking snow \\
  \end{tabular}
  \label{table:best_and_worst_concept_pairs}
\end{table}

\vspace{-0.1in}

%%%%%%%%%%%%%%%%%%%%%%%%%%%%%%%%%%%%%%%%%%%%%%%%%%%%%%%%%%%%%%
%
%	Result Section
%
%
%
%%%%%%%%%%%%%%%%%%%%%%%%%%%%%%%%%%%%%%%%%%%%%%%%%%%%%%%%%%%%%%
\section{Results and Discussion}

In this section we present and analyze the results of the two main experiments, detection and multi-class classification. Within these experiments we discuss the consistency between the pairs and the sounds. As well as the best and worst performing pairs, the most confusable pairs and performance of pairs based on shared adjectives, verbs or nouns. 

The overall detection performance of ANPs and VNPs was better than expected. As a reference, similar experiments with AudioSet~\cite{hershey2017cnn} had comparable performance (AUC 85\%). AudioSet has 485 sound classes and did not deal with the subjectivity treated in this paper. The performance for both types of pairs, ANs and VNs, is shown in Table~\ref{table:detection_results} with the following results: accuracy of 69\% and 71\%, f-score of 51\%, and 53\% and AUC of 70\% and 72\%, respectively. %Because there is no similar work in the literature, we compared our results with random performance, which corresponds to 0.13\% for ANPs and 0.27\% for VNPs. 
The detection results correspond to the extracted audio features of 13 MFCCs, while 20 MFCCs were not included because they did not provide a performance gain. One explanation why VNPs performed better is because verbs tend to be less subjective or more neutral than adjectives~\cite{Neviarouskaya2009}. This means that the acoustic characteristics may be more distinguishable for classifiers as it is for humans. For example, people might argue about what would be the sound of a \textit{beautiful car}, but not so much about the sound of a \textit{passing car}.

We show the best and worst performing ANPs and VNPs in Table~\ref{table:best_and_worst_concept_pairs}. The best detected ANPs had over 93\% accuracy and corresponded to distinguishable sounds of phone numbers being pressed or phones' tones, such as \textit{industrial phone}, \textit{echoing phone} and \textit{weird cell}. On the contrary, \textit{noisy glitch} and \textit{extreme noise} had accuracy around 1\% and corresponded to pairs with adjectives that described a generic meaning rather than specific.  %BME Check last sentence with Sebastian
The best detected VNPs were \textit{howling dog}, \textit{howling wolf}, \textit{crying insects} and \textit{howling animal} with accuracy greater than 94\%. These sounds tend to have almost no overlapping audio. On the contrary, \textit{splashing water}, \textit{crackling footsteps}, \textit{splashing river}, \textit{gurgling water} and \textit{breaking snow} had accuracy around 2\%. These wide-band, background-noise-like continuous sounds are hard to classify. A similar problematic arose in~\cite{UrbanSound} with sounds such as \textit{air conditioning} and \textit{engine idling}. %BME Check last sentence with Sebastian, ALSO check the plausibility score
Additionally, pairs corresponding to long duration recordings, commonly related to environmental sounds and field recordings, tend to have lower performances. %Labels such as noise tend to be  \textit{noise}

%detection, adjective sentence is made up, accuracy is made up
We looked at the overall performance of pairs sharing a common adjective, verb or noun to estimate how well we could detect them. The adjective with the highest accuracy, 71\%, corresponded to \textit{industrial}, which commonly paired with nouns such as \textit{hands}, \textit{phone} and \textit{metal}. On the other hand, the verb with the highest accuracy, 73\%, corresponded to \textit{singing}, which commonly paired with \textit{choir}, \textit{crowd}, \textit{man}, \textit{child} and \textit{woman}. %Interestingly, \textit{singing voice} did not yield a high performance mainly because it was commonly confused with man, woman and child. 
Nouns such as \textit{alert}, \textit{phone} and \textit{guitar} performed well for different adjectives and verbs because they have a specific timbre. %In contrast, examples of nouns with lower performance are \textit{power}, \textit{glitch} and \textit{action}.  %add accuracy

%%%% MULTICLASS  %%%%
The overall multi-class classification of ANPs and VNPs yielded good performance. The accuracies shown in Table~\ref{table:multi_class_classification_results} are respectively for RF and CNN, 1.6\% and 2.1\% for ANPs and 4.5\% and 7.4\% for VNPs. While the multi-class performance is lower than the detection experiments, it is still higher than random performance, which corresponds to 0.13\% for ANPs and 0.27\% for VNPs. As supported in the literature~\cite{piczak2015environmental}, the CNN outperformed the RF algorithm. Similar to the detection case, the RF numbers correspond to audio features of 13 MFCCs. The performance difference between ANPs and VNPs may be explained because there are more than twice number of ANPs than there are VNPs, which makes the classification task harder for ANPs because the classifier has to discern between more pair-classes. 

Moreover, an issue with multi-class setup is that some audio files corresponded to more than one pair label. When a classifier is trained sharing one or more audio files for one or more classes, it struggles to define a decision boundary to separate the classes. A solution could be to add more (different) training audio files, which could ameliorate the problem by aiding the classifier to generalize the class boundaries. Nevertheless, this is an issue that has to be further explored because it is expected that sounds can be labeled with more than one adjective, verb or noun.

The multi-class setup also allowed us to observe pairs that were commonly confused by their acoustic characteristics as shown in Table~\ref{table:interesting_confusions}. The confusion matrix is not included here for lack of space, but in general, the confusions look conceptually reasonable for both types of pairs. For the ANPs, confusions happened when ANPs shared the same adjective and when pairs shared similar contexts expressed by the noun, such as in \textit{extreme rain} and \textit{heavy thunder}. A similar case was observed with VNPs, where the confusions happened when the verb and the noun expressed similar meanings, such as\textit{ splashing lake} and \textit{walking river}.
%Some ANP pairs have a dominant user and the results support the intuition that performance will be high. The dominance of the user translates into having (different) audio files from the same user or group of users on training, CV and testing. Hence, the classifier tend to learn other acoustic characteristics related to the recording procedures. None of the pairs selected for best or worst correspond to user dominated pairs. % Check with Sebastian this last sentence 

These experiments and results evidence a degree of consistency between tag-pairs and the sounds associated to them, despite of the setup limitations and assumptions. In this work, we take a first step towards exploiting tag-pair-based actions and properties of sounds automatically from a large-scale repository. We point out some of the challenges, which have not been published to the best of our knowledge, and could be of great use for research if we want to take advantage of the massive amounts of web audio and video recordings. Other lexical combinations could be explored, such as other verb conjugations and adverbs to add meaning to a given action. Also, sequences of lexical combinations may help describe the order actions take place in a specific scene or describe the properties of the acoustic scene. Reliable recognition of the acoustic nuances can benefit several applications. For example, VNPs can support violent detection in image processing~\cite{datta2002person}. In another example, ANPs could be use for opinion mining to determine the quality of urban soundscapes~\cite{guastavino2006ideal}. Similarly, ANPs could also be combined with the image-based ANPs~\cite{chen2014deepsentibank} for sentiment analysis of videos. Hence, we encourage further exploration of the nuances as audio-only recordings and as a complement of similar research on images and text in multimedia analysis.

\begin{table}[t]
  \centering
  \caption{Overall multi-class classification performance for Random Forest and CNN.}
\begin{tabular}{c|c|c|c}
    Classifier & Pair & \textbf{Features} & \textbf{Accuracy\%} \\  \hline  
	\multirow{1}{*} RF &{ANP} & 13 MFCCs+$\Delta$+$\Delta\Delta$ & 1.6\\   	
    \multirow{1}{*} RF &{VNP} & 13 MFCCs+$\Delta$+$\Delta\Delta$ & 4.5 \\\hline
    \multirow{1}{*} CNN &{ANP} & 13 MFCCs+$\Delta$+$\Delta\Delta$ & 2.1\\   	
    \multirow{1}{*} CNN &{VNP} & 13 MFCCs+$\Delta$+$\Delta\Delta$ & 7.4 \\
  \end{tabular}
  \label{table:multi_class_classification_results}
\end{table}

\begin{table}[t]
  \centering
    \caption{Each row correspond to an example of pairs that were highly confused.}
  \begin{tabular}{c c}
    \multicolumn{2}{c}{\textbf{ANP}} \\
    \hline
    extreme rain & heavy thunder \\
    heavy thunder & heavy wind \\
    distant rain & distant thunder \\
    relaxing water & relaxing creek \\
    echoing church bell & echoing hall \\
    \\

    \multicolumn{2}{c}{\textbf{VNP}} \\
    \hline
    passing railway & passing train \\
    singing bird & tweeting bird \\
    talking crowd & walking noise \\
    splashing lake & waving river \\
    burning fire & crackling fire \\
  \end{tabular}
  \label{table:interesting_confusions}
\end{table} 

%\subsection{Limitations and Future Work}

%Moreover, this issue is not unique to our task, but it is also present in other major audio concept datasets such as the YouTube-based AudioSet~\cite{AudioSet}.

%There were cases where and adjective could also mean a verb and thus creating ambiguity of meaning. The issue happened because the tags and therefore pairs did not follow a word order. Nevertheless, we include the pair in both types. 
%For example, \textit{relaxing band} and \textit{band relaxing} could mean a soft music coming from a band or the act of a band chatting and  

%%%%%%%%%%%%%%%%%%%%%%%%%%%%%%%%%%%%%%%%%%%%%%%%%%%%%%%%%%%%%%
%
%	Conclusion
%
%
%
%%%%%%%%%%%%%%%%%%%%%%%%%%%%%%%%%%%%%%%%%%%%%%%%%%%%%%%%%%%%%%
\vspace{-0.1in}

\section{Conclusions}
In previous years, sound recognition has been used to identify sounds also called audio concepts. However, sounds have nuances that may be better described by adjective-noun pairs such as \textit{breaking glass}, and verb-noun pairs such as \textit{calm waves}. Datasets with these types of labels are unavailable. In this work we provide an investigation about the relation between audio content and weak labels corresponding to Adjective-Noun Pairs and Verb-Noun Pairs. For this study, we collected, processed and made available, AudioPairBank, a large-scale corpus consisting of 761 ANPs and 362 VNPs corresponding to over 33,000 audio files. Using this dataset, we evaluated classification performance of audio recordings and the results supported a degree of consistency between tag-pairs and sounds. We provided a benchmark better than random performance, regardless of complications such as using weak labels and the tagging assumptions from a collaborative repository. We expect to guide researchers on several fronts. One is to provide direction on which pairs to consider for manual annotations and to encourage the exploration of more lexical combinations of sounds. Another front is to further research that analyze sentiment, affect and opinion mining in audio. Lastly, is to show the extent in which audio pairs could be used to complement similar pairs with other modalities.  

\ifCLASSOPTIONcaptionsoff
  \newpage
\fi



\bibliographystyle{IEEEtran}
\bibliography{IEEEfull}
\end{document}